\documentclass[prl,aps,amssymb,showpacs,twocolumn]{revtex4}
\usepackage{amsmath}
\usepackage{amssymb}
\usepackage{amsthm}
\usepackage{amsfonts}
\usepackage{enumerate}
\usepackage{latexsym}
\usepackage[dvips]{graphicx}

\newcommand{\beq}{\begin{equation}}
\newcommand{\eneq}{\end{equation}}

\input{epsf}

\begin{document}

\tolerance 10000

\newcommand{\vk}{{\bf k}}


\title{ On the Nature of Spin Currents }
\author { B. A. Bernevig}

\affiliation{Department of Physics, Stanford University,
         Stanford, California 94305 }

\begin{abstract}
Full expressions for finite frequency spin, charge conductivity and
spin susceptibility in Rashba and Luttinger-type systems are given.
Whereas in the Rashba Hamiltonian the spin conductivity has the same
frequency dependence as the dielectric polarizability \cite{rashba}
and magnetic susceptibility \cite{erlingsson, dimitrova}, the
Luttinger case is different. Moreover, for a generalized Rashba-type
coupling the three quantities also exhibit different frequency
dependencies.
\end{abstract}


\maketitle

In this short note we provide the full frequency dependence of spin
conductivity, dielectric function, and magnetic susceptibility in
Rashba and Luttinger-type systems. In general, for the Luttinger
case or for a generalized Rashba-type spin-orbit coupling, the
frequency dependencies are different, but for the two dimensional
Rashba Hamiltonian they become the same, as noticed before
\cite{rashba, erlingsson, dimitrova}. However, this seems to be the
exception rather than the rule.

Recent theoretical work predicts the existence of spin currents in
semiconductors with spin-orbit coupling placed under the influence
of an electric field \cite{murakamiscience, sinova}. In one of the
proposals \cite{murakamiscience}, spin current is induced in
p-doped cubic bulk semiconductors with Luttinger-type spin-orbit
coupling of the spin-$3/2$ valence band when an electric field is
applied to the material. In the other proposal \cite{sinova} spin
current is induced in an n-doped 2-dimensional semiconductor layer
upon the application of an in-plane electric field. In both cases,
the electric field, the spin polarization and the spin direction
of flow are mutually perpendicular. The spin currents are
'dissiplationless' in the sense that they do not depend on the
momentum scattering rate, as normal charge current does. The
presence of a charge current due to the applied electric field
does however cause dissipation.

Apart from the experimental proof of their existence (the idea of an
experimental spin-voltmeter is yet unimaginable, but experimental
detection of spin accumulation due to spin current at the boundaries
of a sample has recently been reported \cite{wunderlich}), many
other fundamental questions on the nature of spin currents remain.
In a recent paper \cite{rashba}, Rashba focuses on the essential
physical question of whether spin current is a stand-alone phenomena
or is it described macroscopically through the spin-orbit
contribution to the dielectric function. He finds that for the case
of the Rashba Hamiltonian the frequency dependence of the spin
conductance and dielectric function is the same, strongly pointing
to the fact that the two are indeed the same effect. Also,
Erlingsson \emph{et al} and Dimitrova relate the spin current in the
Rashba model to the magnetic susceptibility, linking it to the
uniform spin polarization that also develops in the presence of an
applied electric field. As we show below, this is not the case in
the Luttinger Hamiltonian, nor is it the case for a generalized
Rashba-type coupling.

Let us start with a generalized conduction band Rashba-type
spin-orbit Hamiltonian:
\begin{equation} \label{spinonehalf}
H_R = \varepsilon (k) + \lambda_i ({\bf {k}}) \sigma_i, \;\;\;
\varepsilon(k) = \frac{ k^2}{2m_e}, \;\;\; i = 1,2,3
\end{equation}
\noindent where $m_e$ is the effective mass in the conduction band,
$\sigma_i$ are the Pauli matrices and $\lambda_i({\bf {k}})$ are
generic functions of the momentum ${\bf k}$ which respect the
symmetry of the underlying crystal (the Rashba Hamiltonian is given
by the particular case of a 2 dimensional $k$ space and by
$\lambda_x = \alpha k_y, \; \lambda_y = - \alpha k_x, \;
\lambda_3=0$ but in general we can have all three $\lambda_i$
non-zero such as in the bulk Dresselhaus splitting.) The energy
eigenvalues are $E_\pm = \varepsilon(k) \pm \lambda (k)$ where
$\lambda (k) =\sqrt{\lambda^i \lambda^i}$. The Green's function
reads:
\begin{equation}
G(\textbf{k}, i \omega) = \frac{1}{ i\omega + \mu -H_R} = \frac{
i\omega + \mu - \varepsilon(k) + \lambda_i(k) \sigma_i}{(i\omega +
\mu - \varepsilon(k))^2 - \lambda^2(k)}
\end{equation}
\noindent and the charge current operator is $J_i = \frac{\partial
H_R}{\partial k_i}$ where $i= 1,2,3$. The spin-orbit induced part
of the polarisability tensor $\epsilon_{ij}$ is related to the
charge conductivity tensor $\sigma_{ij}$ and to the response
function $Q_{ij}$ in the following way:
\begin{multline} \label{rashbachargeconductance}
\\
\epsilon_{ij}(\omega) = \frac{4 \pi i \sigma_{ij}
(\omega)}{\omega} \\ \\
\sigma_{ij}(\omega) = \frac{e^2}{2 \hbar} \frac{Q_{ij}(\omega)}{-i \omega} \\ \\
Q_{ij}(i\nu_m)= - \frac{1}{V} \int_0^{\beta} \langle T J_i(u)
J_j\rangle
e^{i \nu_m u} du = \\
=\frac{1}{V \beta} \sum_{k,n} \mathrm{tr}(J_i G(\textbf{k},
i(\omega_n +
\nu_m)) J_j G(\textbf{k}, i \omega_n)) = \\
 =- \frac{2}{V} \sum_k
\frac{n^F(E_+) -n^F(E_-)}{\lambda [ ( i\nu_m)^2 - (2 \lambda)^2]}
\times
\\ \times [\nu_m \epsilon_{abc} \lambda_b \frac{\partial \lambda_a}{\partial
k_i}\frac{\partial \lambda_c}{\partial k_j}+ 2 \lambda^2
(\frac{\partial \lambda}{\partial k_i}\frac{\partial
\lambda}{\partial k_j} - \frac{\partial \lambda_a}{\partial
k_i}\frac{\partial \lambda_a}{\partial k_j})]
\end{multline}
\noindent where $\nu_m = 2 \pi /\beta$, $n^F$ represents the Fermi
function, $\epsilon_{abc}$ is the totally antisymmetric tensor and
we sum on any repeated index. As an essential observation, the
current response function of Eq.[\ref{rashbachargeconductance}]
does \textbf{not} depend on the kinetic energy $\varepsilon(k)$
(except through the Fermi functions) and is entirely given by the
spin-orbit coupling terms $\lambda_i$.

Let us now compute the spin-current charge-current correlation
function $Q^l_{ij}$ which gives the response of the spin current
$J^l_i$ to an applied electric field $E_j$. The spin current
operator for spin flowing in the $i$ direction polarized in the
$l$ direction is $J^l_i = \frac{1}{2} \{ \frac{\partial
H}{\partial k_i}, \sigma^l \}$. The spin current in this system is
\textbf{not} a conserved quantity. The spin conductivity
$\sigma^l_{ij}$ and the spin current-charge current correlation
function read $Q^l_{ij}$ read:
\begin{multline} \label{rashbaspincurrent}
\\
\sigma^l_{ij} = e \frac{ Q^l_{ij}(\omega)}{-i
\omega} \\ \\
   Q^{l}_{ij} (i \nu_m) = -\frac{1}{V} \int_0^{\beta}
\langle T J^{l}_i(u) J_j \rangle  e^{i \nu_m u} du =
 \\
=\frac{1}{V \beta} \sum_{k,n} \mathrm{tr}(J^l_i G(\textbf{k},
i(\omega_n +
\nu_m)) J_j G(\textbf{k}, i \omega_n)) = \\
=-\frac{2}{V } \sum_{k} \frac{n^F(E_+ ) - n^F(E_-)}{\lambda((i \nu_m)^2 - (2\lambda)^2)} \times \\
\times \{\nu_m \frac{\partial \varepsilon}{\partial k_i}
\epsilon_{lns} \lambda_n \frac{\partial \lambda_s}{\partial k_j} + 2
\lambda \frac{\partial \varepsilon}{\partial k_i} (\lambda^l
\frac{\partial \lambda}{\partial k_j} -\lambda \frac{\partial
\lambda_l}{\partial k_j})\}
\end{multline}
\noindent In a marked difference from the dielectric function in
Eq.[\ref{rashbachargeconductance}], the spin conductivity above
depends on the kinetic energy $\varepsilon(k)$ through the term
$\frac{\partial \varepsilon}{\partial k_i}$. The structure of the
the two terms in Eq.[\ref{rashbachargeconductance}] and
Eq.[\ref{rashbaspincurrent}] is fundamentally different for
generic $\lambda_i(k)$.

For the Rashba Hamiltonian the summation transforms in an integral
over the 2-dimensional k-space and $\lambda_x= \alpha k_y, \;\;
\lambda_y = - \alpha k_x, \;\; \lambda_z= 0, \;\; \lambda = \alpha
k$ and hence the first (reactive) term (which contains the fully
antisymmetric $\epsilon_{abc}$ symbol) in
Eq.[\ref{rashbachargeconductance}] vanishes and we get:
\begin{multline} \label{rashbapolarizability}
\\
\epsilon_{ij} = -\frac{8 \pi e^2}{\hbar \omega^2} \int \frac{d^2
k}{(2 \pi)^2} \frac{n(E_+) - n(E_-)}{\omega^2 - (2 \lambda)^2}
\lambda (\frac{\partial \lambda}{\partial k_i}\frac{\partial
\lambda}{\partial k_j} - \frac{\partial \lambda_a}{\partial
k_i}\frac{\partial
\lambda_a}{\partial k_j}) = \\ \\
=  \delta_{ij} \frac{2  e^2 \alpha^3}{\hbar \omega^2}
\int_{k_+}^{k_-}
\frac{k^2 dk}{(2 \alpha k)^2- \omega^2} \\
\end{multline}
\noindent where $k_\pm$ are the Fermi momenta of the $E_\pm$
bands. The above expression is in agreement with \cite{rashba}.
The spin conductance in Eq.[\ref{rashbaspincurrent}] sees the last
term vanish and gives:
\begin{multline}
\\ \sigma^l_{ij} (\omega) =  \frac{ Q^l_{ij} (\omega)}{-i \omega} =
\\ \\
=-\frac{2 e}{ V } \sum_{k} \frac{n(E_+ ) - n(E_-)}{ \lambda
({\omega^2 - (2 \lambda)^2})} \cdot \frac{\partial
\varepsilon}{\partial k_i} \epsilon_{lnm} \lambda_n \frac{\partial
\lambda_m}{\partial k_j}
\end{multline}
\noindent  The only nonzero value is for $l = 3$ (spin polarized
out of plane with in-plane electric field):
\begin{equation} \label{rashbaspinconductivity}
\sigma^3_{12} =  \frac{ e \alpha}{2 \pi  m} \int_{k_+}^{k_-}
\frac{k^2 dk}{(2 \alpha k)^2- \omega^2}
\end{equation}
\noindent As noticed before \cite{rashba}, the spin-orbit part of
the dielectric function and the spin conductivity in
Eq.[\ref{rashbapolarizability}] and
Eq.[\ref{rashbaspinconductivity}] have the same frequency
dependence. However, this is due to $\lambda_i$ being linear in the
momentum components $k_j$. This is the case in pure Rashba and
Dresselhauss systems, as well as for a generic linear combination of
the two. For $\lambda$ linear in $k$, both factors $\frac{\partial
\varepsilon}{\partial k_i} \epsilon_{lns} \lambda_n \frac{\partial
\lambda_s}{\partial k_j}$ from the spin conductance in Eq[4] and
$\lambda^2 (\frac{\partial \lambda}{\partial k_i}\frac{\partial
\lambda}{\partial k_j} - \frac{\partial \lambda_a}{\partial
k_i}\frac{\partial \lambda_a}{\partial k_j})$ from the
polarizability in Eq[3] are quadratic in $k$ since $\partial
\lambda_i /\partial k_j$ is a constant and $\varepsilon$, as kinetic
energy, is quadratic in $k$. A generic $\lambda$ would give
different, spectrum-specific dependence.

The magnetic susceptibility in the Rashba model exhibits similar
behavior. For generic spin-orbit coupling $\lambda$ we have:
\begin{multline}
\chi_{ij}(i\nu_m)= - \frac{1}{V} \int_0^{\beta} \langle T
\sigma_i(u) \sigma_j\rangle
e^{i \nu_m u} du = \\
=\frac{1}{V \beta} \sum_{k,n} \mathrm{tr}(\sigma_i G(\textbf{k},
i(\omega_n +
\nu_m)) \sigma_j G(\textbf{k}, i \omega_n)) = \\
 =- \frac{2}{V} \sum_k
\frac{n^F(E_+) -n^F(E_-)}{\lambda [ ( i\nu_m)^2 - (2 \lambda)^2]}
[\nu_m \epsilon_{ijk} \lambda_k + 2 (\lambda^2 \delta_{ij} -
\lambda_i \lambda_j)]
\end{multline}
\noindent and we see that this differs in tensor structure from both
the spin conductance and the polarizability, both of which contain
derivatives of the spin-orbit coupling. However, for the Rashba
Hamiltonian we obtain:
\begin{equation}
\chi_{ij} (\omega) = - \frac{\alpha}{\pi} \int_{k_+}^{k_-} \frac{
k^2 dk }{  (2 \alpha k)^2 - \omega^2 }
\end{equation}
\noindent which has the same integral dependence as both the
polarizability and the spin-conductance. This is again because
$\lambda$ is linear in $k$. However, a relation between the spin
current and the magnetic susceptibility, (but not them being
identical, as in the Rashba case), should exist in the generic case
on general grounds. In spin $1/2$ systems spin orbit coupling can
always be thought of as a fictitious $k$-dependent, internal
magnetic field (the spin orbit coupling term is always first order
in the spin operator $\lambda_i(k) \sigma_i$ since for spin $1/2$
products of spin operators can be always expressed as linear
combination of the 3 Pauli matrices). The presence of an electric
field gives an non-zero expectation value for the momentum $<k>$ and
hence for the fictitious internal magnetic field $<\lambda (k)>$.
This creates a uniform magnetization proportional and the
coefficient of proportionality is $\chi$. However, since spin is not
conserved in spin $1/2$ systems, it will precess around the internal
magnetic field $\lambda(k)$. This gives rise to an extra-term in the
continuity equations which turns out to be proportional to the spin
current, as in \cite{erlingsson}.  For the Rashba case, the two
effects are related, as shown in \cite{erlingsson} and this relates
the spin conductance to the susceptibility, as also shown above.

We now turn our attention to the Luttinger Hamiltonian for spin
$S=3/2$ holes in the valence band of centrosymmetric cubic
semiconductors:
\begin{equation}
H_L = \frac{1}{2m} (\gamma_1 + \frac{5}{2} \gamma_2) k^2 - \frac{
\gamma_2}{m} (\mathbf{k} \cdot \mathbf{S})^2
\end{equation}
\noindent It can be cast in a form similar to the Rashba case by
the transformation \cite{murakamiprb}:
\begin{multline}
\Gamma^1 = \frac{2}{\sqrt 3} \{S_y, S_z \}, \;\; \Gamma^2 =
\frac{2}{\sqrt 3} \{S_z, S_x \}, \;\; \Gamma^3 = \frac{2}{\sqrt 3}
\{S_y, S_x \}\\
 \Gamma^4= \frac{1}{\sqrt 3} (S_x^2 - S_y^2), \;\;
\Gamma^5 = S_z^2 - \frac{5}{4} I_{4 \times 4},
\end{multline}
\noindent which satisfy the $SO(5)$ Clifford algebra $\Gamma^a
\Gamma^b + \Gamma^b \Gamma^a = 2 \delta_{ab} I_{4 \times 4}$. The
Hamiltonian becomes:
\begin{equation}
H_L=\varepsilon (k) + d_a\Gamma^a, \;\; \varepsilon (k)  =
\frac{\gamma_1}{2m} k^2, \;\; a=1,...,5 \label{Clifford}
\end{equation}
\noindent where
\begin{equation}
\begin{aligned}
 \\ d_1 & = - \sqrt{3}
\frac{\gamma_2}{m} k_z k_y, \; d_2 = - \sqrt{3} \frac{\gamma_2}{m}
k_x k_z, \;
d_3 = - \sqrt{3} \frac{\gamma_2}{m} k_x k_y,\\
d_4 &= - \frac{\sqrt{3}}{2} \frac{\gamma_2}{m} (k_x^2 - k_y^2), \;
d_5 =- \frac{1}{2} \frac{\gamma_2}{m} (2k_z^2 - k_x^2 - k_y^2)
\end{aligned}
\end{equation}
where $\gamma_1, \gamma_2$ are material dependent parameters, and
$m$ is the electron mass. Since it is a time-invariant parity even
fermionic hamiltonian, the band structure is composed of two doubly
degenerate bands called light and heavy hole bands corresponding to
helicity $\pm 1/2$ and $\pm 3/2$ with energies $E_\pm =
\varepsilon(k) \pm d(k)$, $d^2=d_ad_a$. The Green's function reads:
\begin{equation}
G(\textbf{k}, i \omega) = \frac{1}{ i\omega + \mu -H_L} = \frac{
i\omega + \mu - \varepsilon(k) + d_a(k) \Gamma_a}{(i\omega + \mu -
\varepsilon(k))^2 - d^2(k)}
\end{equation}
\noindent Introducing as before the current operator $J_i =
\frac{\partial H_L}{\partial k_i}$ we have the following
expression for the spin-orbit coupling part of the dielectric
function:
\begin{multline} \label{luttingerdielectric}
\\
\epsilon_{ij}(\omega) = \frac{4 \pi i \sigma_{ij}
(\omega)}{\omega} \\ \\
\sigma_{ij}(\omega) = \frac{e^2}{2 \hbar} \frac{Q_{ij}(\omega)}{-i \omega} \\ \\
Q_{ij}(i\nu_m)= - \frac{1}{V} \int_0^{\beta} < T J_i(u) J_j>
e^{i \nu_m u} du = \\
=- \frac{8}{V} \sum_k \frac{n^F(E_+) -n^F(E_-)}{ ( i\nu_m)^2 - (2
d)^2} d (\frac{\partial d}{\partial k_i}\frac{\partial d}{\partial
k_j} - \frac{\partial d_a}{\partial k_i}\frac{\partial
d_a}{\partial k_j})]
\end{multline}
\noindent Using the identity $\frac{\partial d_a}{\partial
k_i}\frac{\partial d_a}{\partial k_j} = (\frac{\gamma_2}{m})^2 (
k_i k_j + 3 k^2 \delta_{ij})$, as well as doing the integrals over
the angles we get
\begin{multline} \label{dielectrick} \\
\epsilon_{ij} (\omega)  =  \delta_{ij}  \frac{16  e^2
\gamma_2^3}{\pi \hbar m^3 \omega^2} \int_{k_+}^{k_-}
\frac{k^6 dk}{(2 \frac{\gamma_2}{m} k^2)^2- \omega^2} \\
\end{multline}
\noindent The spin conductivity is obtained as the response of the
spin current to an applied electric field. A straight-forward but
tedious calculation gives:
\begin{multline} \label{spinconductivity}
\\ \sigma^l_{ij} (\omega) = e \frac{ Q^l_{ij} (\omega)}{-i \omega}
\\ \\
Q^l_{ij} (i \nu_m)  =-\frac{4  \nu_m}{ V } \sum_{k}
\frac{n^F(E_+ ) - n^F(E_-)}{d ( i \nu_m)^2 - (2 d)^2} \times \\ \\
\times \eta^l_{ab} \left[2 d_b \frac{\partial d_a}{\partial k_j}
\frac{\partial \varepsilon}{\partial k_i} + \epsilon_{abcde} d_e
\frac{\partial d_c}{\partial k_i} \frac{\partial d_d}{\partial
k_j} \right]
\end{multline}
\noindent where $Q^{l}_{ij} (i \nu_m) = -\frac{1}{V}
\int_0^{\beta} \langle T J^{l}_i(u) J_j \rangle  e^{i \nu_m u} du$
and where $\eta^{l}_{ab}$, $l=1,2,3$, $a,b =1,..., 5$ is a tensor
antisymmetric in $a,b$ relating the spin-$3/2$ matrix $S^l$ to the
$SO(5)$ generators $\Gamma^{ab} = \frac{1}{2i} [\Gamma^a,
\Gamma^b]$, $S^l = \eta^l_{ab} \Gamma^{ab}$. The explicit form of
$\eta^{l}_{ab}$ is \cite{murakamiprb}:
\begin{multline}
\\ \eta^1_{15} =\frac{\sqrt{3}}{4}, \;\; \eta^1_{23} =-\frac{1}{4},
\;\; \eta^1_{14} =\frac{1}{4}, \;\; \eta^2_{25} =
-\frac{\sqrt{3}}{4}, \;\; \\ \\
 \eta^2_{13} = \frac{1}{4}, \;\;
\eta^2_{24} =\frac{1}{4}, \;\; \eta^3_{34} =-\frac{1}{2} , \;\;
\eta^3_{12} = -\frac{1}{4}
\end{multline}
 From
Eq.[\ref{luttingerdielectric}] and Eq.[\ref{spinconductivity}] we
can see that the spin current and dielectric function again have
very different structure. This is reflected when we introduce the
explicit $k$ dependence:
\begin{equation} \label{spinconductivityk}
\sigma^k_{ij} (\omega)  =  \epsilon_{ijk} \frac{e \gamma_2}{\pi^2
m^2}(\gamma_1 +\frac{2 \gamma_2}{ 3} )\int_{k_+}^{k_-} \frac{k^4
dk}{(2 \frac{\gamma_2}{m} k^2)^2- \omega^2}
\end{equation}
\noindent From Eq.[\ref{dielectrick}] and
Eq.[\ref{spinconductivityk}] we see that they have different
spectra. The result for spin conductivity in this note differs from
the result in \cite{murakamiprb} because we use the full spin
operator while \cite{murakamiprb} use a conserved spin operator,
which commutes with the Hamiltonian. Different from the case of the
Rashba Hamiltonian, here we can define a conserved spin current by
projecting the spin-$3/2$ operator into the light and heavy-hole
bands \cite{murakamiprb}. The expression for spin conductivity in
this case takes an extremely nice, topological form and can be
obtained in our case by neglecting the first term in the square
bracket of Eq.[\ref{spinconductivity}] or by neglecting the
$\gamma_1$ term in Eq.[\ref{spinconductivityk}]. The conserved spin
exists in the Luttinger Hamiltonian due to the presence of
degenerate sub-bands, and it is rigourous only in inversion
symmetric semiconductors where the bands are doubly degenerate.
Different from the spin $1/2$ case (Rashba, Dresselhauss, etc.)
where the spin is not conserved and precesses around a fictitious
internal magnetic field given by the spin-orbit coupling, the
conserved spin will not precess and will satisfy a continuity
equation. While in general spin is defined as the physical quantity
which couples to a magnetic field, in this case the spin-$3/2$ in
the valence band of cubic semiconductors is actually a combination
of both angular spin-$1$ and Pauli spin $1/2$ degrees of freedom.
Hence it is not immediately obvious whether the spin or the
conserved spin is the meaningful physical quantity which would
couple to an external field. Depending on whether the gap between
the heavy and light hole states at the Fermi energy is larger than
thermal and disorder energies or not it might be the case that
either the conserved or the non-conserved spin is the physically
measurable quantity.

We now turn to the calculation of magnetic susceptibility in the
Luttinger model. We start with the usual, non-conserved spin and
find:
\begin{multline} \label{spinconductivity}
\chi_{ij} (i \nu_m) = - \frac{1}{V} \int_0^{\beta} < T S_i(u) S_j>
e^{i \nu_m u} du  =\\ = \frac{4  }{ V } \delta_{ij} \sum_{k}
\frac{n^F(E_+ ) - n^F(E_-)}{ d( i \nu_m)^2 - (2 d)^2} \eta^i_{ab}
\eta^j_{bc} d_a d_c
 \\ = \frac{4  }{ V } \sum_{k}
\frac{n^F(E_+ ) - n^F(E_-)}{ ( i \nu_m)^2 - (2 d)^2} d = \\
=- \delta_{ij} \frac{2}{\pi^2} \frac{\gamma_2}{m} \int_{k_+}^{k_-}
\frac{k^4 dk}{(2 \frac{\gamma_2}{m} k^2)^2- \omega^2}
\end{multline}
\noindent where $\omega = i \nu_m$. We see that this is has the same
functional form (frequency dependence) as the spin conductance,
thought not the same tensor structure. Since we have computed the
magnetic susceptibility of the usual, non-conserved spin, this is
expected, by arguments similar to the spin $1/2$ case, since the
non-conserve spin also precesses. However, when we compute the spin
conductance of the conserved spin $S_i^{cons} = P_{HH} S_i P_{HH} +
P_{LH} S_i P_{LH}$ where $P_{HH}$ and $P_{LH}$ are the projection
operators in the heavy and light holes, we obtain:
\begin{equation}
\chi_{ij} (i \nu_m) = - \frac{1}{V} \int_0^{\beta} < T S^{cons}_i(u)
S^{cons}_j> e^{i \nu_m u} du  = 0
\end{equation}
\noindent The magnetic susceptibility vanishes whereas the spin
conductance is finite.

We now attempt to make some general statements about spin currents
in spin-orbit coupling systems. First consider spin-1/2 coupling
systems given by the generic Hamiltonian in Eq.[\ref{spinonehalf}].
Since they have no degenerate states, in these systems there is no
conserved spin. If the spin-orbit coupling is linear in $k$
($\lambda \sim k$) then the spin conductivity is 'universal' in the
sense that it does not depend on the spin-orbit coupling strength.
This is valid for both 2 and 3 dimensions, as long as the Fermi
energy is positive. The universal spin conductance is an artifact of
the spin-orbit coupling being linear in the momentum.

For spin 3/2 systems with time-reversal and parity invariance one
can define a conserved spin current with finite response to an
electric field by projecting the spin onto the degenerate states.
The nice topological form obtained in \cite{murakamiprb} is entirely
due to the fact that the system is spin-$3/2$. Spin $3/2$ systems do
not have universal conductance as the the Hamiltonian is quadratic
and not linear in $k$.

The spin current conductivities computed here are valid for the
disorder-free case. The introduction of disorder in the Rashba case
gives the remarkable result that the vertex correction completely
cancels the disorder-free spin current \cite{inoue}. This does not
happen in the Luttinger case due to the inversion symmetry of the
Luttinger Hamiltonian. Moreover tedious calculations reveal that
vertex correction does not cancel the spin conductivity in a variety
of other Hamiltonians, including $k^3$ Dresselhauss and Rashba plus
$k$-linear Dresselhauss spin splitting. In this sense, it seems that
the cancelation in the Rashba case is accidental and not due to a
symmetry (Ward identity) of the system. 

In the present short note we have computed the dielectric function,
the spin conductivity and the magnetic susceptibility for both
Rashba and Luttinger type Hamiltonians in a way which makes their
gauge structure clear, and showed that, in general, they have
different frequency dependencies. We also discussed some generic
issues about spin-orbit coupling systems.

We wish to thank S.C. Zhang for countless discussions on the subject
of spin-orbit coupling and spin currents. Support was offered
through the Stanford SGF program as well as by the NSF under grant
numbers DMR-0342832 and the US Department of Energy, Office of Basic
Energy Sciences under contract DE-AC03-76SF00515.

\end{document}